\documentclass{article}
\usepackage{amscd,amsmath,amssymb}
\usepackage[english]{babel}

\def\beq{\begin{equation}}
\def\eeq{\end{equation}}
\def\ba{\beq\begin{array}{c}}
\def\ea{\end{array}\eeq}

\def\p{\partial}

\def\e{{\epsilon}}

\def\cO{\mathcal{O}}

\begin{document}

\begin{titlepage}
  \centerline{}
  \vskip 2in
  \begin{center}
    \large{\Large{\bf On the Magnetohydrodynamics/Gravity Correspondence}}
    \vskip 0.5in
    Vyacheslav Lysov
    \vskip 0.3in
    {\it Center for the Fundamental Laws of Nature,
      Harvard University\\
      Cambridge, MA, 02138}
  \end{center}
  \vskip 0.5in
  \begin{abstract}
    The fluid/gravity correspondence relates solutions of the
    incompressible Navier-Stokes equation to metrics which solve the
    Einstein equations. In this paper we extend this duality to a new
    magnetohydrodynamics/gravity correspondence, which translates
    solutions of the equations of magnetohydrodynamics (describing
    charged fluids) into geometries that satisfy the Einstein-Maxwell
    equations. We present an explicit example of this new correspondence
    in the context of flat Minkowski space. We show that a
    perturbative deformation of the Rindler wedge satisfies the
    Einstein-Maxwell equations provided that the parameters appearing
    in the expansion, which we interpret as fluid fields, satisfy the magnetohydrodynamics equations. As a
    byproduct of our analysis we show that in four dimensions, the
    dual geometry is algebraically special Petrov type II.
  \end{abstract}
\end{titlepage}

\setcounter{page}{1}
\pagenumbering{arabic}

\tableofcontents

\section {Introduction}

In the last fifteen years, the holographic viewpoint has become
increasingly central to our understanding of black hole physics. An
explicit duality between fluid dynamics and black hole geometries has
been established by several groups
\cite{Bhattacharyya:2008kq,Bredberg:2011jq,Huang:2011he,Compere:2011dx,
  Eling:2012ni}. Some constructions make use of the hydrodynamic
expansion, while others rely on the near-horizon expansion or special
algebraic properties of the metric. In each case, regardless of the
details, the Navier-Stokes equation always captures the low-energy
horizon dynamics. We may regard this as the universal aspect of the
correspondence. However, some of its details do depend on the
particular setting in which the correspondence is applied. For
instance, the dual fluid may obey the relativistic Navier-Stokes
equation \cite{Compere:2012mt,Eling:2012ni,Bhattacharyya:2008ji}, or
it may be subject to particular forcing terms
\cite{Bhattacharyya:2008kq,Bredberg:2011jq}.

When faced with the full complexity of the nonlinear Navier-Stokes
equation, one may be tempted to start looking for a solution in the
absence of external forces. However, there is a particular choice of
forcing term arising from the Lorentz force which has been extensively
studied in plasma physics. This suggests that it may be advantageous
to establish a new version of the correspondence for the charged
fluid, which would allow us to bring our knowledge of the
magnetohydrodynamics (MHD) to bear on the problem of solving the
Einstein-Maxwell equations. Previous attempts to establish such a
correspondence \cite{Zhang:2012uy} have considered background
magnetic fields interacting with the fluid. Instead, we propose to
examine dynamical magnetic fields (induced by the fluid's motion)
without background electromagnetic fields.

As such, we work in the Rindler wedge of flat Minkowski space and
investigate perturbations of the geometry in the hydrodynamic limit,
subject to some boundary conditions. Our perturbation, which we carry
out to third order, is parametrized by fluid fields which satisfy the
MHD equations. Interestingly, we find that in this setup, the
conductivity  $\sigma$ of the charged fluid is precisely the reciprocal of its
fluid viscosity $ \eta =1/4\pi \sigma $. Similarly to \cite{Bredberg:2011jq}, we show that the
dual metric, after some suitable rescaling, depends on only one
parameter, which is obtained from a combination of the derivative
expansion parameter and the distance between the metric horizon and
the fluid surface. It is therefore possible to translate the
derivative expansion into the near-horizon expansion, and vice versa.

In some cases \cite{Bredberg:2011jq} the fluid/gravity duality
involves metrics which are algebraically special. In four dimensions, we
show that the metric dual to MHD obeys Petrov type II conditions up to
third order.

The main results of the paper are in section 5, where  we
formulate the Cauchy problem for the Einstein-Maxwell equations, and
describe one of its solutions in a hydrodynamic expansion. This
analysis is preceded by a short review of the Einstein-Maxwell theory
(section 2) and of the MHD equations (section 3) and their scaling
properties (section 4). In section 6, we elaborate on our solution by
constructing it order by order in perturbation theory. We supplement this presentation
with sections 7 and 8, in which we provide some additional
details about the near-horizon expansion and offer some checks for the
Petrov type II. We conclude the paper with some thoughts, open
questions and possible generalizations.

\section{Einstein-Maxwell theory}

The Einstein-Maxwell equations
\begin{equation}
  G_{\mu\nu}=8\pi GT_{\mu\nu}
\end{equation}
describe gravity coupled to the electromagnetic stress tensor
\begin{equation}
  4\pi T_{\mu\nu}=F_{\mu\lambda}{F_\nu}^\lambda-\frac14 g_{\mu\nu}F^2,
\end{equation}
where the gauge field itself solves the Maxwell equations
\begin{equation}
  \nabla_\mu F^{\mu\nu}=0.
\end{equation}
In the remainder of this paper, we work in units where
$8\pi G=1$ and $c=1$. Rather than working directly with the
gauge field $A_\mu$, it is more convenient to use the field strength
$F_{\mu\nu}$ and impose the Bianchi constraint
\begin{equation}
  \nabla_{[\lambda}F_{\mu\nu]}=0.
\end{equation}
These equations are well studied and several exact solutions are
known. A famous example is y the Reissner-Nordstr\"{o}m
solution, which describes a spherically-symmetric charged object in an
asymptotically flat 4-dimensional space. Other known solutions include
planar charged objects obeying Anti-deSitter asymptotics, as well as
less familiar gravitational-wave-like solutions with both metric and
gauge field fluctuations. We will also be interested in the latter
type of solution.

\section{Magnetohydrodynamics}

In $3+1$ dimensions, the MHD equations with finite conductivity
$\sigma$ take the following form \cite{LL8}:
\begin{gather}
  \nabla\cdot B=0,\qquad\nabla\cdot v=0,\nonumber\\
  \nabla\times E=-\p_tB,\nonumber\\
  \nabla\times B=4\pi J=4\pi\sigma(E+[v\times B]),\\
  \p_tv+(v\cdot\nabla)v+\nabla P-\eta\nabla^2v=[J\times B].\nonumber
\end{gather}
The equations can be partially solved for the electric field $E$ and
current $J$. The remaining equations then form a nonlinear system
describing a fluid with velocity $v$, subject to pressure $P$ and a
magnetic field $B$:
\begin{equation}
\begin{gathered}
  \nabla\cdot B=0,\qquad\nabla\cdot v=0,\\
  \p_tB=\nabla\times[v\times B]+\frac{1}{4\pi\sigma}\nabla^2B,\\
  \p_tv+(v\cdot\nabla)v+\nabla P-\eta\nabla^2v
  =\frac{1}{4\pi}(B\cdot\nabla)B-\frac{1}{8\pi}\nabla\!\left(B^2\right).
\end{gathered}\label{vecmh}
\end{equation}
As in the case of the incompressible Navier-Stokes equation, there are
sufficiently many equations to determine all the variables. It will
prove useful to rewrite the system (\ref{vecmh}) in terms of the electromagnetic
fields $f_{ij}$, $f_{\tau i}$, for $i=1,\ldots,3$ 
($E_i = f_{i\tau},\;\; B_i = \frac12 \e_{ijk}f^{jk}$)
\begin{equation}\label{mhd}
\begin{gathered}
\p_{[k}f_{ij]}=0,\qquad\p_iv^i=0,\\ 
  \p_\tau f_{ij}=\p_if_{\tau j}-\p_jf_{\tau i},\qquad
  f_{\tau i}=-\frac{1}{4\pi\sigma}\p_jf_{ij}-v^kf_{ki},\\
  \p_\tau v_i+v^j\p_jv_i+\p_iP-\eta\p^2v_i+\p^j\pi_{ji}=0,\qquad
  \pi_{jk}=\frac{1}{4\pi}\left(f_{jl}f_{kl}-\frac14 f^2\delta_{jk}\right).
  \end{gathered}
\end{equation}

The MHD equations (\ref{mhd}) can be generalized to higher dimensions
by assuming $i=1,\ldots,p$.

\section{Scaling properties}

The Navier-Stokes equation (equation (\ref{mhd}) with no
electromagnetic field) is famous for its scaling property:
simultaneous rescaling of the coordinates and fields
\begin{align}
  v(x,\tau)\quad&\longrightarrow\quad
  \e v\!\left(\e x,\e^2\tau\right),\\
  P(x,\tau)\quad&\longrightarrow\quad
  \e^2P\!\left(\e x,\e^2\tau\right),\notag
\end{align}
leaves the equation invariant while preserving the viscosity
$\eta$. This scaling property is responsible for the universality of
the NS equation in capturing the low energy dynamics of fluids.
It may be extended to the MHD equations (\ref{mhd}) by requiring that
the electromagnetic field obey the following scaling relation:
\begin{align}
  f_{ij}(x,\tau)&\quad\longrightarrow\quad
  \e f_{ij}\!\left(\e x,\e^2\tau\right)\\
  f_{i\tau}(x,\tau)&\quad\longrightarrow\quad
  \e^2 f_{i\tau}\!\left(\e x,\e^2\tau\right).\notag
\end{align}
The scaling properties of the MHD equations allow us to write an
ansatz for the bulk gauge field:
\begin{align}
  F_{ij}&=\e F_{ij}^0+\e^2F^1_{ij}+\ldots\\
  F_{i\tau}&=\e^2 F_{ir}^0+\e^3F_{i\tau}^1+\ldots\notag
\end{align}
In order to ensure that the Bianchi identities hold at each expansion
order independently (so that fields of different orders do not mix),
the rest of the components should be chosen to be:
\begin{align}
  \label{irrt}
  F_{ir}&=\e^0F_{ir}^0+\e F_{ir}^1+\ldots\\
  F_{r\tau}&=\e^1F_{r\tau}^0+\e^2F_{r\tau}^1+\ldots\notag
\end{align}

\section {MHD/gravity correspondence}

We can now describe the MHD/gravity correspondence in the simplest
possible setup. The starting point is the flat Minkowski metric in
$(p+2)$-dimensional space,
\begin{equation}
  ds^2=-rd\tau^2+2d\tau dr+dx_i^2,\qquad i=1,\ldots,p,
\end{equation}
with no background electromagnetic field. The hyper surface $\Sigma_c$
at fixed radius $r=r_c$, whose induced metric is flat, is the
background space in which the fluid theory evolves\footnote{The Brown-York
  stress tensor is diagonal and can be trivially identified with the
  fluid stress tensor at rest.}. We will now study a perturbative
deformation of the metric and electromagnetic field which obeys the
Einstein-Maxwell equations
\begin{gather}
  G_{\mu\nu}=2G\left(F_{\mu\lambda}{F_\nu}^\lambda-\frac14 g_{\mu\nu}F^2\right),\\
  \nabla_\mu F^{\mu\nu}=0,\qquad
  \nabla_{[\lambda}F_{\mu\nu]}=0,\qquad\mu=r,\tau,1,\ldots,p,
\end{gather}
and the following boundary conditions:

{\bf Regularity at the horizon:} both the field strength $F$ and the
metric are regular at $r=0$.

{\bf Dirichlet boundary conditions:} the induced metric on $\Sigma_c$
is a flat Minkowski metric, and there is no induced charge nor current
on $\Sigma_c$, i.e.
\begin{equation}
  \label{embound}
  F^{ir}(r_c)=F^{\tau r}(r_c)=0.
\end{equation}
One of the solutions to the Cauchy problem is\footnote{Details of the
  derivation are provided in the next section.}
\begin{align}
  \label{dualmetric}
  ds_{p+2}^2=&-rd\tau^2+2d\tau dr+dx_idx^i
  -2\left(1-\frac{r}{r_c}\right)v_idx^id\tau-2\frac{v_i}{r_c}dx^idr\notag\\
  &+\left(1-\frac{r}{r_c}\right)
  \left[(v^2+2P)d\tau^2+\frac{v_iv_j}{r_c}dx^idx^j\right]
  +\left(\frac{v^2}{r_c}+\frac{2P}{r_c}\right)d\tau dr\notag\\
  &-\frac1{16\pi p}{\left(1-\frac{r}{r_c}\right)^2}f^2d\tau^2
  +\frac1{2\pi r_c}\left(1-\frac{r}{r_c}\right)
  \left(f_{ik}f_{jl}\delta^{kl}-\frac14 \delta_{ij}f^2\right)dx^idx^j\\
  &-\frac{\left(r^2-r_c^2\right)}{r_c}\p^2v_idx^id\tau
  +\cO\left(\e^3\right),\notag\\
  r_cF=&\,\frac12 f_{ij}dx^i\wedge dx^j+f_{i\tau}dx^i\wedge d\tau
  -\p_jf_{ij}dx^i\wedge dr+\cO\left(\e^3\right).\notag
\end{align}
This geometry, which is parametrized by the fluid fields $f_{ij}$,
$v_i$, $P$ and $f_{i\tau}$ (which depend only on $x^i$ and $\tau$),
will satisfy the Einstein-Maxwell equations to order $\cO(\e^4)$
provided that the fluid fields satisfy the MHD equations,
\begin{gather}
  \p_\tau v_i+v^j\p_jv_i+\p_i\left(P-\frac{p+2}{16\pi p}f^2\right)-r_c\p^2v_i+\p^j\pi_{ji}=0,\qquad
  \pi_{jk}=\frac{1}{4\pi}\left(f_{jl}f_{kl}-\frac14 f^2\delta_{jk}\right),
  \notag\\
  \p_iv^i=0,\notag\\
  \p_\tau f_{ij}=\p_if_{\tau j}-\p_jf_{\tau i},\\
  \p_{[k}f_{ij]}=0,\notag\\
  f_{\tau i}=-r_c\p_jf_{ij}-v^kf_{ki}.\notag
\end{gather}
Interestingly, $\pi_{ij}$ is the lowest component of the
electromagnetic energy momentum tensor on $\Sigma_c$ in the
$\e$-expansion.  Moreover, the two diffusion constants which enter the
MHD equations turn out to be equal:
\begin{equation}
  \eta=\frac{1}{4\pi\sigma}=r_c.
\end{equation}
Perhaps this relation is not unexpected for such a simple background
metric, since there are no dimensionless parameters for this ratio to
depend on.

\section{Solution}

The Cauchy problem described in section 5 is generally hard to solve exactly. Nevertheless, due to the scaling properties of the MHD system,
it is possible to construct a perturbative solution,  as was done in \cite{Bredberg:2011jq}. The expansion assumes small perturbation size and slowly-varying spacetime dependence:
\begin{equation}
  v_i\sim\cO(\e),\quad\p_i\sim\cO(\e),\quad
  \p_\tau\sim\cO(\e^2),\quad P\sim\cO(\e^2),\quad
  f_{ij}\sim\cO(\e),\quad f_{i\tau}\sim\cO(\e^2).
\end{equation}
The problem may be simplified even further, as follows. At each given
order in the expansion, we may divide the equations into two groups:
constraint equations and propagating equations. The former depend only
on the data from lower orders because of the extra spatial $\p_i$ and
time derivatives $\p_\tau$ present, whereas the latter fix the radial
dependence of the new metric components introduced at the same
order. The Navier-Stokes equation with magnetic forcing is a
constraint equation which appears at third order $\cO(\e^3)$ so it can
be written in terms of the metric solution at $\cO(\e^2)$ order.

In the remainder of this section, we will construct the geometry up to
and including the $\cO(\e^2)$ order and describe the constraint
equations at $\cO(\e^3)$ order.

\subsection{Zeroth order }

The background metric is flat Minkowski space, which solves Einstein's equations with no source terms:
\begin{equation}
  ds^2=-rd\tau^2+2d\tau dr+dx_idx^i.
\end{equation}
At zeroth order $\cO(\e^0)$, there is one nontrivial Maxwell equation:
\begin{equation}
  \p_r\left(rF^{0}_{ri}\right)=0.
\end{equation}
The only solution that is regular at $r=0$ is the trivial solution
$F_{ri}^0=0$.

To summarize, at zeroth order $\cO(\e^0)$, the solution is:
\begin{gather}
  ds^2=-rd\tau^2+2d\tau dr+dx_idx^i+\cO(\e),\\
  F=\cO(\e).\notag
\end{gather}
         
\subsection{First order}

Next, we wish to introduce a deformation of the metric parameterized
by the fluid fields $v^i$ and $P$. The simplest way to do this is to
use small Lorentz boosts, as was done in \cite{bkls}, resulting in
\begin{equation}
  ds^2=-rd\tau^2+2d\tau dr+dx_idx^i
  -2\left(1-\frac{r}{r_c}\right)v_idx^id\tau
  -2\frac{v_i}{r_c}dx^idr+\cO(\e^2).
\end{equation}
There are no corrections to the electromagnetic field since the
background field vanished in the first place. At first order in $\e$,
the Maxwell equations are:
\begin{align}
  \p_rF_{\tau r}^0=0\qquad&\Longrightarrow\qquad
  r_cF_{\tau r}^0=Q^0(x,\tau),\notag\\
  \p_r\left(rF_{ri}^1\right)=0\qquad&\Longrightarrow\qquad
  F^1_{ri}=0,\\
  \p_rF_{ij}^0=0\qquad&\Longrightarrow\qquad
  r_cF_{ij}^0=f_{ij}(x,\tau).\notag
\end{align}
In the above, $Q^0$ can be interpreted as the charge density of the
dual fluid. The only solution satisfies boundary condition
(\ref{embound}) corresponds to $Q^0=0$.

In summary, the solution to first order $\cO(\e^1)$ is:
\begin{gather}
  ds^2=-rd\tau^2+2d\tau dr+dx_idx^i
  -2\left(1-\frac{r}{r_c}\right)v_idx^id\tau
  -2\frac{v_i}{r_c}dx^idr+\cO\left(\e^2\right),\\
  r_cF=\frac12f_{ij}dx^i\wedge dx^j+\cO\left(\e^2\right).\notag
\end{gather}

\subsection{Second order}

Note that the zeroth and first order solutions did not impose any
constraints on the fluid fields $v^i$ and $f_{ij}$. On the other hand,
in order to solve the second order equations, we will need to impose
some constraints on the fluid fields. In addition, we will have to
introduce some extra fields such as $P(x,\tau)$ and $ f_{\tau i}(x,\tau)$.

At second order $\cO(\e^2)$, the Maxwell equations are:
\begin{align}
  \p_rF_{i\tau}^0=0\qquad&\Longrightarrow\qquad
  r_cF_{i\tau}^0=f_{i\tau}(x,\tau),\notag\\
  \p_{[k}F_{ij]}^0=0\qquad&\Longrightarrow\qquad
  \p_{[k}f_{ij]}=0,\notag\\
  \p_rF_{ij}^1=0\qquad&\Longrightarrow\qquad
  r_cF_{ij}^1=f^1_{ij}(x,\tau),\\
  \p_rF_{\tau r}^1=0\qquad&\Longrightarrow\qquad
  F_{\tau r}^1=Q^1(x,\tau),\notag\\
  \p_r\left(rF^2_{ri}+F^0_{\tau i}\right)+\p_jF^0_{ji}=0
  \qquad&\Longrightarrow\qquad
  r_cF_{ri}^2=\p_jf_{ij}.\notag
\end{align}
Here, $Q^1$ is the next order correction to the charge density. In
order to satisfy the boundary conditions at $\Sigma_c$, we must set
$Q^1=0$ and also require that
\begin{equation}
  F^{ri}(r_c)=0\qquad\Longrightarrow\qquad r_cF_{ri}^2+F_{\tau i}^0+v^kF_{ki}=0,
\end{equation}
which has the following solution:
\begin{equation}
  f_{\tau i}=-r_c\p_jf_{ij}-v^kf_{ki}. 
\end{equation}
The equation above is one of the MHD equations (\ref{mhd}) that we
obtained by solving the Einstein-Maxwell equations at second order
$\cO(\e^2)$. Having obtained the field strength to this order, we can
evaluate the stress tensor to second order as well:
\begin{gather}
  r_c^2F^2=f^2+\cO\left(\e^3\right),\qquad
  4\pi r_c^2T_{ij}=f_{li}f_{lj}-\frac14 f^2\delta_{ij},\notag\\
  4\pi r_c^2T_{r\tau}=-\frac14 f^2,\qquad
  4\pi r_c^2T_{\tau\tau}=\frac{r}{4}f^2,\\
  4\pi T_{rr}=\cO\left(\e^3\right),\qquad
  4\pi T_{\tau i}=\cO\left(\e^3\right),\qquad
  4\pi T_{ri}=\cO\left(\e^3\right).\notag
\end{gather}
The nontrivial contributions to the stress tensor will backreact on
the metric and produce additional terms of order $\cO(\e^2)$ in
$g_{ij}^{(2)}$ and $g_{\tau\tau}^{(2)}$. For example, the $(i,j)$ component of
the Einstein equations will take the form
\begin{equation}
  R_{ij}=-\frac12 \p_r\left(r\p_rg_{ij}^{(2)}\right)
  =8\pi G\left(T_{ij}-\frac{1}{p}T^\mu_{\mu}\delta_{ij}\right).
\end{equation}
Using the boundary condition on $\Sigma_c$, we can write the solution
in the form
\begin{align}
  g_{ij}^{(2)}&=\frac{1}{2\pi r_c}\left(1-\frac{r}{r_c}\right)
  \left(f_{li}f_{lj}-\frac1{2p}f^2\delta_{ij}\right),\\
  g^{(2)}_{\tau\tau}&=-\frac{2+p}{16 \pi p}\left(1-\frac{r}{r_c}\right)^2f^2.\notag
\end{align}
As aforementioned, at this order in the expansion, we must introduce a
constraint equation. In this case, it amounts to the requirement that
the velocity field be divergence free:
\begin{equation}
  \p_i v^i=0.  
\end{equation}
To summarize, at second order $\cO(\e^2)$, the solution is defined by:
\begin{align}
  ds_{p+2}^2=&-rd\tau^2+2d\tau dr+dx_idx^i 
  -2\left(1-\frac{r}{r_c}\right)v_idx^id\tau-2\frac{v_i}{r_c}dx^idr\notag\\
  &+\left(1-\frac{r}{r_c}\right)
  \left[\left(v^2+2P\right)d\tau^2+\frac{v_iv_j}{r_c}dx^idx^j\right]
  +\left(\frac{v^2}{r_c}+\frac{2P}{r_c}\right)d\tau dr\notag\\
  &+\frac{1}{2\pi r_c}\left(1-\frac{r}{r_c}\right)
  \left(f_{li}f_{lj}-\frac1{2p}f^2\delta_{ij}\right)dx^idx^j
  -\frac{2+p}{16 \pi p}\left(1-\frac{r}{r_c}\right)^2f^2d\tau^2+\cO(\e^3),\\
  r_cF=&\,\frac12 f_{ij}dx^i\wedge dx^j+\frac12 f^1_{ij}dx^i\wedge dx^j
  +f_{i\tau}dx^i\wedge d\tau-\p_jf_{ij}dx^i\wedge dr+\cO(\e^3),\notag\\
 f_{\tau i}=&\,-r_c\p_jf_{ij}-v^kf_{ki},\qquad\p_{[k}f_{ij]}=0,\notag\\
 \p_iv^i=&\,0.\notag
\end{align}

\subsection{Third order}

As in the second order case, at third order in the $\e$-expansion we must once again introduce new fields and impose additional constraints on the ones that were introduced at lower orders.

To be more precise, the equations which are tangent to $\Sigma_c$ are
constraint equations, while the remaining equations fix the radial
dependence of the geometry at order $\cO(\e^3)$ in terms of the fluid
fields. We illustrate this point in the context of the Bianchi
identity:
\begin{align}
  \p_rF^1_{i\tau}=-\p_iF^1_{\tau r}
  \qquad&\Longrightarrow\qquad
  r_cF^1_{i\tau}=f^1_{i\tau},\notag\\
  \p_rF^2_{ij}=\p_iF^2_{rj}-\p_jF^2_{ri}
  \qquad&\Longrightarrow\qquad
  r_cF_{ij}^2=r(\p_i\p_kf_{jk}-\p_j\p_kf_{jk})+f_{ij}^2,\\
  \p_{[k}F^1_{ij]}=0\qquad&\Longrightarrow\qquad
  \p_{[k}f^1_{ij]}=0,\notag\\
  \p_\tau F^0_{ij}=\p_iF^0_{\tau j}-\p_jF^0_{\tau i}
  \qquad&\Longrightarrow\qquad
  \p_\tau f_{ij}=\p_if_{\tau j}-\p_jf_{\tau i}.\notag
\end{align}
The equations in the last two lines are tangent to $\Sigma_c$ and
therefore impose constraints on the fluid data
$f_{ij},f_{ij}^1,f_{i\tau}$ from lower orders. On the other hand, the
equations in the first two lines fix the $\cO(\e^3)$ radial dependence
of the field strength components in terms of the newly introduced
fluid fields.
 
The Einstein-Maxwell constraint equations on $\Sigma_c$ have the
following form:
\begin{gather}
  n_\mu\nabla_\nu F^{\mu\nu}=0,\notag\\
  n^\nu G_{\mu\nu}=8\pi G\,n^\nu T_{\mu\nu},\\
  n^\nu n^\mu G_{\mu\nu}=8\pi G\,n^\nu n^\mu T_{\mu\nu},\notag
\end{gather}
where $n^\mu$ is a unit normal vector to $\Sigma_c$. At third order
$\cO(\e^3)$, the Maxwell constraint is
\begin{equation}
  \p_i\left(rF^2_{ri}+v^kF^0_{ki}+F^0_{ \tau i}\right)+\p_\tau F^0_{\tau r}=0.
\end{equation}
This condition is trivially satisfied due to our choice of boundary
condition (\ref{embound}). The only nontrivial gravitational
constraint at third order $\cO\left(\e^3\right)$ is:
\begin{equation}
  0=n^\mu G_{\mu i}-8\pi G\,n^\mu T_{\mu i}
  =\frac{1}{2r_c}\left[\p_\tau v_i+v^j\p_jv_i+\p_iP-r_c\p^2v_i+\frac{1}{4\pi}\p^j
    \left(f_{jl}f_{il}-\frac{p+1}{2p}f^2\delta_{ij}\right)\right].
\end{equation}
It can be identified with the last of the MHD equations (\ref{mhd})
after performing the redefinition
\begin{equation}
  P-\left(\frac{p+2}{16\pi p}\right)f^2\quad\longrightarrow\quad P.  
\end{equation}
Having established the solvability of the constraint equations at
$\cO(\e^3)$ order, the Cauchy theorem applied to the Einstein-Maxwell
equations guarantees the existence of a solution for the entirety of
the $\cO(\e^3)$ equations. This full solution differs from
(\ref{dualmetric}) in that it contains additional fluid field terms at
order $\cO(\e^3)$. Finally, note that we may choose $f_{ij}^1=0$,
which trivially satisfies the Bianchi identity. This concludes the
derivation of the duality proposed in section 5.
 
\section{Near-horizon expansion}

In this section, we will establish the equivalence of the
hydrodynamic expansion for the metric (\ref{dualmetric}) to the
near-horizon expansion of the geometry. In order to achieve this, we
begin by performing the coordinate redefinition from
\cite{Bredberg:2011jq}, namely:
\begin{equation}
  x^i=\frac{r_c}{\e}\hat{x}^i,
  \qquad\tau=\frac{r_c}{\e^2}\hat{\tau},
  \qquad r=r_c\hat{r},
\end{equation}
In these new coordinates, the derivatives are no longer assumed to be
small,
i.e. $\hat{\p}_i=\cO(\e^0),\p_{\hat{\tau}}=\cO(\e^0)$, and the dual
metric (\ref{dualmetric}) takes the following form
\begin{align}
  \label{resc}
  \frac{\e^2}{r_c^2}ds_{p+2}^2=&-\frac{\hat{r}}{\lambda}d\hat{\tau}^2
  +2d\hat{\tau}d\hat{r}+d\hat{x}_id\hat{x}^i 
  -2\left(1-\hat{r}\right)\hat{v}_id\hat{x}^id\hat{\tau}  
  +\left(1-\hat{r}\right)\left(\hat{v}^2+2\hat{P}\right)d\hat{\tau}^2
  -\frac1{16\pi p}{\left(1-\hat{r}\right)^2}\hat{f}^2d\hat{\tau}^2\notag\\
  &+\lambda\left[(1-\hat{r})
    \left(\hat{v}_i\hat{v}_j+\frac{1}{2\pi}\hat{f}_{ik}\hat{f}_{jl}\delta^{kl}
      -\frac{1}{4\pi p} \hat{f}^2\delta_{ij}\right)d\hat{x}^id\hat{x}^j
    -2\hat{v}_id\hat{x}^id\hat{r} \right.\\
   &\qquad\qquad\qquad\qquad\qquad\qquad\qquad\left.+\left(\hat{v}^2+2\hat{P}\right)d\hat{\tau}d\hat{r}
    +\left(1-\hat{r}^2\right)
    \hat{\p}^2\hat{v}_id\hat{x}^id\hat{\tau}\right]+\ldots\notag\\
  \frac{\e}{r_c}F=&\,\frac12 \hat{f}_{ij}d\hat{x}^i\wedge d\hat{x}^j
  +\hat{f}_{i\tau}d\hat{x}^i\wedge d\hat{\tau}
  -\lambda\hat{\p}_j\hat{f}_{ij}d\hat{x}^i\wedge d\hat{r}+\ldots\notag
\end{align}
where we introduced a new expansion parameter
$\lambda=\frac{\e^2}{r_c}$ as well as new fluid fields defined by
\begin{align}
  \hat{v}_i(\hat{x},\hat{\tau})
  =\frac{1}{\e}v_i(\hat{x}(x),\hat{\tau}(\tau)),\qquad
  &\hat{P}(\hat{x},\hat{\tau})
  =\frac{1}{\e^2}P(\hat{x}(x),\hat{\tau}(\tau)),\\
  \hat{f}_{ij}(\hat{x},\hat{\tau})
  =\frac{1}{\e}f_{ij}(\hat{x}(x),\hat{\tau}(\tau)),\qquad
  &\hat{f}_{i\tau}(\hat{x},\hat{\tau})
  =\frac{1}{\e^2}f_{i\tau}(\hat{x}(x),\hat{\tau}(\tau)).\notag
\end{align}
After a suitable rescaling, the geometry (\ref{resc}) will no longer
depend on the two independent parameters $r_c$ and $\e$; rather, it
will be parameterized by the single parameter $\lambda$. Likewise, the $r_c$ dependence also drops out of the MHD equations, which become:
\begin{gather}
  \p_{\hat{\tau}}\hat{v}_i+\hat{v}^j\hat{\p}_j\hat{v}_i
  +\hat{\p}_i\left(\hat{P}-\frac{p+2}{16\pi p}\hat{f}^2\right)-\hat{\p}^2\hat{v}_i
  +\frac{1}{4\pi}\hat{\p}^j\left(\hat{f}_{jl}\hat{f}_{il}
    -\frac14 \hat{f}^2\delta_{ij}\right)=0,\\
  \hat{f}_{\tau i}=-\hat{\p}_j\hat{f}_{ij}-\hat{v}^k\hat{f}_{ki}.\notag
\end{gather}
The distance between the metric horizon at $r=0$ and the cutoff
surface at $r=r_c$ in the rescaled metric (\ref{resc}) behaves as
$\frac{1}{\sqrt{r_c}}$, so should not be surprising that there are two
ways to make $\lambda$ small: one way is to perform a hydrodynamic
expansion in $\e\ll 1$ on the fluid surface $\Sigma_c$ while keeping
$r_c$ fixed; the other way consists of pushing the cutoff surface
$\Sigma_c$ close to the horizon ($r_c\gg 1$) while removing the small
derivative restriction on the fluid fields (so that $\e$ can be
arbitrarily large).

\section{Petrov type}

As in \cite{Bredberg:2011jq}, we find that in four dimensions ($p=2$),
the geometry (\ref{dualmetric}) is of algebraically special Petrov
type II, meaning that there exists a null vector $k^\mu$ such that the
Weyl tensor satisfies
\begin{equation}
  W_{\mu\nu\rho[\sigma}k_{\lambda]}k^\nu k^\rho=0.
\end{equation}
One may verify the existence of such a null vector by evaluating the
invariant $I^3-27J^2$, which is a function of the metric. The details
about $I$ and $J$ and their explicit value in terms of the metric
components can be found in \cite{exac}. The lowest nontrivial
components of $I$ and $J$ are typically of order $\cO(\e^4)$ and
$\cO(\e^6)$, respectively. Hence we generally expect the invariant
$I^3-27J^2$ to be of order $\cO(\e^{12})$, while an explicit
computation for the invariant of the metric (\ref{dualmetric}) reveals
it to be of order $\cO(\e^{14})$.
 
\section{Conclusion and open questions}
 
The primary purpose of this work was to show that the fluid/gravity
correspondence can be naturally extended to include electromagnetic
fields, and to shed some light on this new facet of the duality.

We illustrated this new aspect of the correspondence in the simplest
nontrivial background, namely the Rindler wedge of flat Minkowski
space. In that context, we were able to obtain an explicit solution to
the Einstein-Maxwell equations as a hydrodynamic expansion
parameterized by the fluid fields with polynomial bulk dependence. In
the process, we discovered that the dual MHD equations have equal
magnetic and fluid diffusion constants.
 
In light of the results in \cite{Compere:2011dx}, which were cast in a
similar framework to ours \cite{Bredberg:2011jq}, we believe that the
Cauchy problem from section 5 admits a solution at all orders in the
hydrodynamic expansion. In the 4-dimensional case, we were able to
perform a test of the algebraically special character of the geometry,
which turned out to be of Petrov type II. It is very likely that this
statement will continue to hold in higher dimensions, though in such
cases there is no analogue to the invariant $I^3-27J^2$ which can be
used to perform the check. Nevertheless, it should be possible to
generalize our solution to other background geometries. It seems worth
investigating the dimensionless ratio of the two diffusion constants,
as it might be subject to certain restrictions in the case of MHD
theories with gravity duals. In particular, it would be interesting to
find a background corresponding to the infinitely conducting fluid
$\sigma=\infty$, which serves as a good approximation to real world
MHD problems.

In \cite{Bredberg:2011jq}, the observation that the metric was of an
algebraically special type strongly suggested the hypothesis that
algebraically special metrics have fluid duals \cite{Lysov:2011xx}. The fact that the
metric (\ref{dualmetric}) is algebraically special leads us to
formulate a new conjecture: Petrov type I metrics which solve the
Einsten-Maxwell equations with properly aligned electromagnetic field
strength appear to be dual to MHD-like fluid equations on
codimension-one hypersurfaces. In the limit when the mean curvature of
the hypersurface is large, these fluid equations reduce to the usual
MHD equations; some work in this direction was done in
\cite{Zhang:2012uy}.
 
\section{Acknowledgements} 

 I am
indebted to Andrew Strominger for the fruitful discussions we had and the helpful
advice he offered throughout the duration of the project. I am also
grateful to G. Compere, R. Loganayagam, A. Lupsasca and G.-S. Ng for
their illuminating comments. This work was supported by the National Science Foundation 
Award  PHY-1205550.

\bibliography{MHD_ref}{}
\bibliographystyle{utphys}

\end{document}